\documentclass[journal=nalefd]{achemso}

\usepackage{siunitx}
\usepackage{chemformula}
\usepackage{graphicx}
\usepackage{hyperref}

\newcommand{\mr}[1]{\mathrm{#1}}
\newcommand{\ea}[0]{\textit{et al.}}
\newcommand{\cf}[0]{c.f.\ }
\newcommand{\ie}[0]{i.e.\ }
\newcommand{\egg}[0]{e.g.\ }

\newcommand{\fref}[1]{Fig.~\ref{fig:#1}}

\newcommand{\eref}[1]{Eq.~(\ref{eq:#1})}
\newcommand{\Cref}[1]{Chapter~\ref{chap:#1}}
\newcommand{\cref}[1]{Ch.~\ref{chap:#1}}

\renewcommand{\vec}[1]{{\mathrm{\mathbf{#1}}}}

\title{Spin-Hall-Active Platinum Thin Films Grown Via Atomic Layer Deposition}

\author{Richard Schlitz}
\email{richard.schlitz@tu-dresden.de}
\affiliation{Institut f{\"u}r Festk{\"o}rper- und Materialphysik, Technische Universit{\"a}t Dresden, 01062 Dresden, Germany}
\alsoaffiliation{Center for Transport and Devices of Emergent Materials, Technische Universit{\"a}t Dresden, 01062 Dresden, Germany}
\author{Akinwumi Abimbola Amusan}
\affiliation{Leibniz Institute for Solid State and Materials Research Dresden (IFW Dresden), Institute for Metallic Materials, 01069 Dresden, Germany}
\author{Michaela Lammel}
\affiliation{Leibniz Institute for Solid State and Materials Research Dresden (IFW Dresden), Institute for Metallic Materials, 01069 Dresden, Germany}
\alsoaffiliation{Technische Universit{\"a}t Dresden, Institute of Materials Science, 01062 Dresden, Germany}
\author{Stefanie Schlicht}
\affiliation{Friedrich-Alexander University Erlangen-N{\"u}rnberg, Department of Chemistry and Pharmacy, Inorganic Chemistry, 91058 Erlangen, Germany}
\author{Tommi Tynell}
\affiliation{Leibniz Institute for Solid State and Materials Research Dresden (IFW Dresden), Institute for Metallic Materials, 01069 Dresden, Germany}
\author{Julien Bachmann}
\affiliation{Friedrich-Alexander University Erlangen-N{\"u}rnberg, Department of Chemistry and Pharmacy, Inorganic Chemistry, 91058 Erlangen, Germany}
\author{Georg Woltersdorf}
\affiliation{Institute of Physics, Martin Luther University Halle-Wittenberg, 06120 Halle, Germany}
\author{Kornelius Nielsch}
\affiliation{Leibniz Institute for Solid State and Materials Research Dresden (IFW Dresden), Institute for Metallic Materials, 01069 Dresden, Germany}
\alsoaffiliation{Technische Universit{\"a}t Dresden, Institute of Materials Science, 01062 Dresden, Germany}
\author{Sebastian T. B. Goennenwein}
\affiliation{Institut f{\"u}r Festk{\"o}rper- und Materialphysik, Technische Universit{\"a}t Dresden, 01062 Dresden, Germany}
\alsoaffiliation{Center for Transport and Devices of Emergent Materials, Technische Universit{\"a}t Dresden, 01062 Dresden, Germany}
\author{Andy Thomas}
\affiliation{Leibniz Institute for Solid State and Materials Research Dresden (IFW Dresden), Institute for Metallic Materials, 01069 Dresden, Germany}
\alsoaffiliation{Center for Transport and Devices of Emergent Materials, Technische Universit{\"a}t Dresden, 01062 Dresden, Germany}

\begin{document}

\date{\today}

\newpage

\begin{abstract}
	We study the magnetoresistance of yttrium iron garnet/Pt heterostructures
	in which the Pt layer was grown via atomic layer deposition (ALD). Magnetotransport experiments in three orthogonal rotation planes reveal    
	the hallmark features of spin Hall magnetoresistance.
    We estimate the spin transport parameters by comparing the magnitude of the magnetoresistance in samples with different Pt thicknesses. We compare the spin Hall angle and the spin diffusion length of the ALD Pt layers to the values reported for high-quality sputter-deposited Pt films. The spin diffusion length of \SI{1.5}{\nm} agrees well with platinum thin films reported in the literature, whereas the spin Hall magnetoresistance $\Delta \rho / \rho = \SI{2.2e-5}{}$ is approximately a factor of 20 smaller compared to that of our sputter-deposited films. 
	Our results demonstrate that ALD allows fabricating spin-Hall-active Pt films of suitable quality for use in spin transport structures. This work provides the basis to establish conformal ALD coatings for arbitrary surface geometries with spin-Hall-active metals and could lead to 3D spintronic devices in the future.
    
\end{abstract}

\maketitle
Atomic layer deposition (ALD) is a powerful process that allows 3D conformal coatings.\cite{Miikkulainen:2013fm} ALD has been extensively used for the deposition and conformal coating of thin oxide insulator films
onto nanopatterned templates or flat substrates.  Increasingly more metals can also be deposited using ALD, and deposition processes have already been developed for several metals.\cite{Miikkulainen:2013fm, Ramos:2013}

In particular, the ALD of Pt has been investigated by several groups. Different precursor chemistries based on trimethyl(methylcyclopentadienyl)platinum, \ch{Pt(CpMe)Me3},\cite{Aaltonen:2003, Knoops:2008} or platinum acetylacetonate, \ch{Pt(acac)2}, \cite{Haemaelaeinen:2008} have been reported, with the former generally resulting in films with higher conductivity.

Pt with its strong spin-orbit coupling is one of the key materials for modern spintronics, allowing the efficient conversion of charge currents to spin currents and vice versa, i.e., leading to a large spin Hall effect.\cite{Chen:2016, Althammer:2013} Thus, the ALD of Pt could open the door for 3D metallic nanostructures with spintronic functionality, for instance structures dependent on high aspect ratios, such as racetrack memory.\cite{Parkin:2008, Hayashi:2008}

Additionally, interesting phenomena related to spin transport in non-planar geometries (\egg coated nanowires) were recently proposed.\cite{Streubel:2016}

In particular, the propagation length of spin/magnon currents in such curved geometries should crucially depend on the spin current's polarization direction.\cite{Streubel:2016, PhysRevLett.117.227203, PhysRevB.95.184415}

To determine by electrical transport whether spin generation and detection are also feasible in such structures, spin Hall magnetoresistance (SMR) can be used. 
SMR is a powerful tool for determining the 
spin transport parameters in ferromagnetic 
insulator (FMI)/non-ferromagnetic metal (NM) heterostructures.\cite{Chen:2013, Althammer:2013, Nakayama:2013}
In particular, 
the magnitude of the SMR effect as a function of the NM thickness allows inferring the spin Hall angle $\Theta_\mr{SH}$ and the spin diffusion length 
$\lambda_\mr{NM}$ of the normal metal and the FMI/NM interface quality quantified by the spin mixing conductance $G_\mr{r}$.\cite{Chen:2013}

Here, we show that Pt films grown via ALD are indeed spin Hall active. Specifically, we observe an SMR with magnitude \SI{2.2e-5}{} in heterostructures consisting of an yttrium iron garnet (\ch{Y3Fe5O12}, YIG) thin film covered by a Pt layer grown by ALD. 
This is clear evidence for spin-Hall-driven spin current transport across the YIG/Pt interface.

Thus, our study establishes the ALD deposition of Pt, a prototypical material that is widely used as a detector/injector for spin currents. This provides an important contribution toward the realization of spin transport experiments in non-planar/non-trivial geometries and might lead to spintronic applications in 3D geometries in the future.

We started from commercially available,
\SI{1}{\um} thick YIG 
films grown via liquid phase epitaxy on \ch{Gd3Ga5O12} substrates. Then, we
used the established cleaning and pre-preparation procedure to prepare our ex situ YIG/Pt 
samples.\cite{Jungfleisch:2013, Puetter:2017}

\begin{figure}[tbhp] 
    \includegraphics{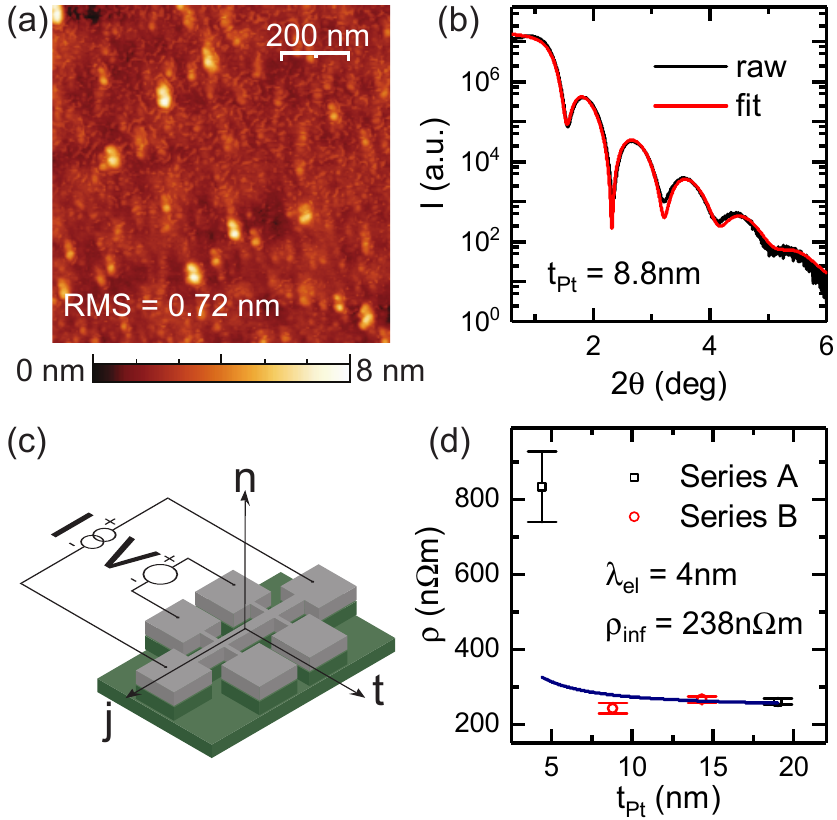}
    \caption{\label{fig:struc}
	Panel (A) depicts an exemplary AFM measurement of a YIG/Pt sample with $t_\mr{Pt} = \SI{8.8}{\nm}$, yielding an rms roughness of $h = \SI{0.72}{\nm}$. 
    An XRR measurement on the same sample and the respective fit are shown in panel (b). 
    Panel (c) displays the sample structure after deposition and lithography. The contacts for the resistivity measurement and the coordinate system with respect to the Hall bar are also depicted here.
    The obtained resistivities for the two sample series are plotted in panel (d) as a function of the platinum thickness. 
    A fit of \eref{fischer} to the data yields an electron mean free path of $\lambda_\mr{el} = \SI{4}{\nm}$ and a 
    bulk platinum resistivity of $\rho_\mr{inf} = \SI{238}{\nano\ohm\meter}$.}
\end{figure}

The YIG films were cleaned using piranha etching solution (3\ch{H2SO4}:1\ch{H2O2}) for one minute to remove organic residue from the surface.\cite{Puetter:2017} 
Subsequently, the samples were submerged in distilled or de-ionized water and loaded into the ALD chamber while still  covered with water. The two different sets of equipment and parameters that were used for growing the Pt films are denoted as series A and series B.

The growth of series A was performed in a commercially available GemStar XT-R thermal bench-top ALD system from Arradiance. \ch{Me(CpPt)Me3} was used as the Pt precursor with pure oxygen (\ch{O2}) as the oxidizer. The chamber temperature was set to \SI{250}{\celsius}, and the organic precursor was preheated to \SI{68}{\celsius} in order to increase the evaporation rate. The pulse and exposure times of the \ch{Me(CpPt)Me3} were set to \SI{50}{\milli\second} and \SI{20}{\second}, respectively, followed by a \SI{60}{\second} pumping time for the removal of any residual precursor and the reactants. For pulsing the Pt precursor, the so-called boost mode was used, in which \ch{Ar} was inserted into the \ch{Me(CpPt)Me3} precursor bottle to increase the amount of precursor inserted into the chamber. For the second half-cycle, \ch{O2} was pulsed for \SI{20}{\milli\second} with subsequent exposure and pumping times of \SI{4}{\second} and \SI{60}{\second}, respectively. For the samples grown within series A, 100 and 280 cycles were performed, resulting in thicknesses of $t_\mathrm{Pt}=(4.4 \pm 1)\,$nm and $t_\mathrm{Pt}=(19.0 \pm 1)\,$nm, respectively.

The platinum films for series B were grown in a Gemstar-6 ALD reactor, which is also commercially available from Arradiance. The same organic precursor was used, but the oxidizer was replaced with ozone (\ch{O3}) due to its higher reactivity. The ozone was provided by a BMT 803N ozone generator. The organic precursor was heated to \SI{50}{\celsius}, while the reactor chamber was set to \SI{220}{\celsius}. The pulse and exposure times of the Pt precursor were set to \SI{500}{\milli\second} and \SI{30}{\second}, respectively. The pulse and exposure steps were performed two times to ensure a saturation of the sample surface with the organic precursor. Afterward, the precursor residue and the reactants were purged from the chamber in a \SI{90}{\second} pump interval. Subsequently, \ch{O3} was pulsed for \SI{500}{\milli\second}, followed by an exposure time of \SI{30}{\second} and a pump time of \SI{90}{\second}. Herewith, 160 (240) cycles result in a Pt thickness of $t_\mathrm{Pt}=(8.8 \pm 0.5)\,$nm ($t_\mathrm{Pt}=(14.3 \pm 0.5)\,$nm). A summary of the growth parameters of both series A and B is presented in Tab. \ref{tab:para}. 

\begin{table}
\begin{tabular}{|c||c|c|}
	\hline 
	\rule[-1ex]{0pt}{2.5ex}  & series A & series B \\ 
	\hline
	\rule[-1ex]{0pt}{2.5ex} Chamber temperature T$_{ch}$ [$^{\circ}$C] & 250 & 220 \\ 
	\hline 
	\rule[-1ex]{0pt}{2.5ex} Precursor temperature T$_{Pt}$ [$^{\circ}$C] & 68 & 50 \\ 
	\hline 
	\rule[-1ex]{0pt}{2.5ex} Pt: t$_{p}$/t$_{exp}$/t$_{pump}$ [s] & 0.05*/20/60 & 0.5/30$^{\dagger}$/90 \\ 
	\hline 
	\rule[-1ex]{0pt}{2.5ex} O$_2$\textbar O$_3$: t$_{p}$/t$_{exp}$/t$_{pump}$ [s] & 0.02/4/60 & 0.5/30/90 \\ 
	\hline 
\end{tabular} 
	\caption{The growth parameters used in series A and B are summarized in this table. The process flow is defined by pulse times (t$_{p}$), exposure times (t$_{exp}$) and pump times (t$_{pump}$). The precursor flow was increased using N$_2$ for the pulse time marked by an asterisk *. The steps before the dagger $^{\dagger}$ were performed twice before continuing with the process.}
\label{tab:para}
\end{table}

Additionally, a reference sample was prepared with a \SI{7}{\nm} sputtered Pt film, where the YIG film was 
additionally annealed in the ultra-high-vacuum of the deposition chamber at 
\SI{200}{\celsius} after the piranha etch to further
improve the interfacial quality and to mimic the temperature of the ALD process. 

To investigate the surface topology, atomic force microscopy (AFM) was performed to extract the rms roughness of our films.
An AFM measurement of a sample with $t_\mr{Pt}=\SI{8.8}{\nm}$ yields a roughness of \SI{0.72}{\nm} that is consistent with comparable films shown in the literature.\cite{Althammer:2013}
Furthermore, the exact thickness of the films was determined by X-ray reflectometry (XRR) measurements and subsequent fitting of the obtained curves. An exemplary set of data and the respective fit are shown in \fref{struc}(b).

After finishing the structural characterization, Hall bars were patterned into the Pt layers (cf. \fref{struc}(a))
using optical lithography and subsequent dry etching with \ch{Ar} ions. 
The Hall bars have a length of $l = \SI{400}{\um}$ and a width of $w = \SI{80}{\um}$. 

To establish electrical contact to our setup, the samples were glued to a chip carrier and contacted via wedge bonding with aluminum wire.
To quantify the magnetoresistive response, the samples were mounted in a magnet setup with a cylindrical Halbach array.\cite{Halbach:1980} It features a constant magnetic flux density of $\mu_0 H = \SI{1}{\tesla}$ perpendicular to the array's cylindrical axis.

To obtain the magnetoresistance, we drive a current of $I = $\SIrange{80}{500}{\uA} along
the Hall bar with a Keithley 2450 sourcemeter while
simultaneously recording the voltage drop with a Keithley 2182 nanovoltmeter. 
To further improve the measurement sensitivity
and to remove spurious contributions, we employ a current reversal technique.\cite{Goennenwein:2015} 

The resistivity of the samples as a function of the platinum thickness is shown in \fref{struc}(d). As expected for platinum and all other metals, a sharp increase in the resistivity toward low thicknesses is observed, which is consistent with previous reports.\cite{Vlietstra:2013, Meyer:2014} We use \eref{fischer} to fit the data and extract the mean free path in our platinum layers assuming that we are in the diffusive limit.\cite{Fischer:1980} The fit yields a bulk resistivity of $\rho_\mr{inf}=\SI{238}{\nano\ohm\meter}$ and an electron mean free path of $\lambda_\mr{el}=\SI{4}{\nm}$ when using the roughness of $h=\SI{0.72}{\nm}$ as determined by AFM.

\begin{equation}
	\rho(t_\mr{Pt}) = \rho_\mr{inf} \left(1+\frac{3 \lambda_\mr{el}}{8(t_\mr{Pt}-h)}\right)
	\label{eq:fischer}
\end{equation}
The extracted mean free electron path and the bulk resistivity agree well with values reported for evaporated platinum thin films.\cite{Meyer:2014} 

To determine the angular dependence of the magnetoresistance, the Halbach array and thus the magnetic field are rotated around the cylindrical axis. Using three different sample inserts, we define the (mutually orthogonal) rotation planes of the magnetic field.
For in-plane rotations (ip), the magnetic field is rotated in the film plane around the surface normal $\vec{n}$. For the other two rotation planes, with a finite component of the magnetic field out of the film plane (oop), the magnetic field is either rotated around the direction of the current flow $\vec{j}$ (oopj) or the transverse direction $\vec{t}$ (oopt). The three rotation planes are shown as insets in \fref{admr}(a-c).


\begin{figure}[th]
    \includegraphics{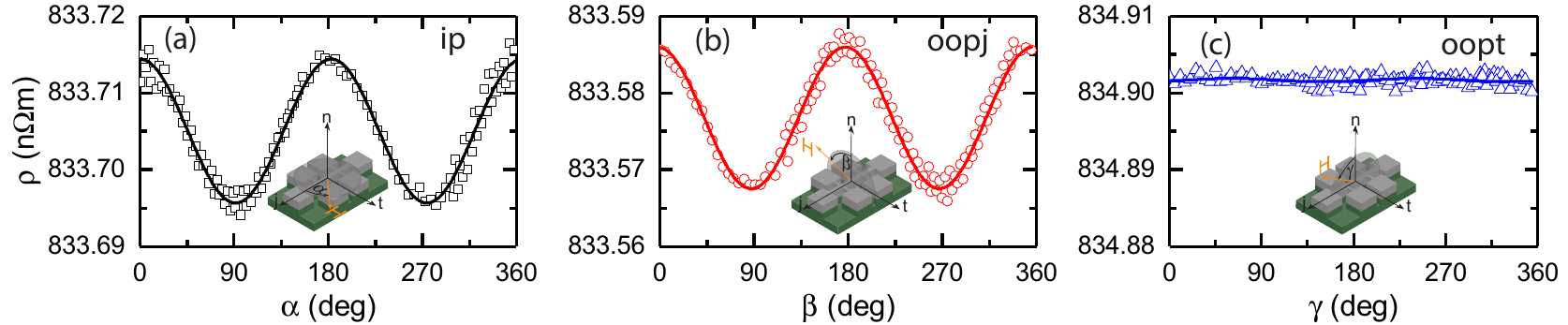}
    \caption{\label{fig:admr} 
    The resistance of a YIG/Pt sample with $t_\mr{Pt}=\SI{4.4}{\nm}$ obtained during
    rotations of the magnetic field in ip, 
    oopj and oopt configurations is shown in
    (a), (b) and (c), respectively. The definitions of the three orthogonal rotation planes ip, oopj and oopt are shown as insets in the respective panels. All data were collected at room temperature
    with a constant magnetic flux density $\mu_0 H = \SI{1}{\tesla}$. A linear slope was subtracted from the data. 
    A $\sin^2(\alpha, \beta)$ modulation of $\rho$ is evident only for the ip and oopj rotations, indicating
    the presence of a magnetoresistance, having a symmetry consistent with spin Hall magnetoresistance.}
\end{figure}

The obtained magnetoresistance for a YIG/Pt (ALD) film with $t_\mr{Pt} = \SI{4.4}{\nm}$
is shown in \fref{admr}. The resistivity of Pt is strongly temperature dependent; therefore, a linear drift was subtracted from the data to compensate for the slow drifts of the sample temperature. Since the SMR only depends on the projection of the magnetization onto the $\vec{t}$ direction,\cite{Althammer:2013, Chen:2013} \ie, $\rho \propto m_\vec{t}^2$, we expect to observe a $\sin^2(\alpha, \beta)$ modulation for the ip and oopj configurations and no modulation for the oopt rotation. This is fully corroborated by our experimental observations. In other words, \fref{admr} shows the characteristic fingerprint of SMR in our YIG/Pt heterostructures also for ALD-grown Pt.

The magnitude of the SMR for the sample shown in \fref{admr} is $\Delta \rho/\rho = \SI{2.2e-5}{}$.
Comparing these values to our reference sample with a sputtered Pt 
film ($\Delta \rho / \rho = \SI{3.6e-4}{}$), the SMR amplitude is reduced by a factor of 20 and is smaller by a factor of 40 when compared to the best YIG/Pt heterostructures with similar Pt thicknesses.\cite{Althammer:2013} This result leads to two possible conclusions: either the interface of the heterostructure is not ideal or the quality of the Pt film is decreased by using ALD. However, the electrical characterization of our films contradicts the latter. Consequently, we assume that contributions such as organic contaminants at the interface or the cleaning procedure should be further optimized to take ALD-specific requirements into account.

To further analyze the relevant transport parameters in our heterostructures, we investigate the thickness dependence of the SMR (\cf \fref{mrdpt}). As expected for SMR, the magnitude of the MR decreases for increasing thickness.\cite{Chen:2013} 
\begin{equation}
	\frac{\Delta \rho}{\rho} = \frac{2 \Theta_\mr{SH}^2 \lambda_\mr{Pt}^2 \rho_\mr{Pt} G_\mr{r}}{ t_\mr{Pt}} \frac{\tanh^2(\frac{t_\mr{Pt}}{2\lambda_\mr{Pt}})}{1 + 2 \rho_\mr{Pt} \lambda_\mr{Pt} G_\mr{r} \coth(\frac{t_\mr{Pt}}{\lambda_\mr{Pt}})}
	\label{eq:smrdpt}
\end{equation}

Using two different sets of parameters adapted from the study by Althammer \ea\cite{Althammer:2013} together with the bulk resistivity $\rho = \SI{238}{\nano\ohm\meter}$ and \eref{smrdpt}, we can reproduce the trend of the thickness dependence (\cf dark red and dark blue curves in \fref{mrdpt}). 
However, to also obtain a good fit of the magnitude of our data, we have to reduce the spin mixing conductance by approximately a factor of 10. The two sets of parameters are summarized in \fref{mrdpt}. Additionally, from the two parameter sets, it is clear that the spin Hall angle and the spin mixing conductance are closely related and that their influence cannot be trivially separated. Nevertheless, all parameters agree well with the range of previously reported values.\cite{Chen:2016}

In the ALD grown samples, the interface quality is most likely affected by the organic constituents of the precursor\cite{Puetter:2017}, making novel approaches in the pre-treatment of the YIG films prior to deposition necessary. 

\begin{figure}[tbhp]
    \includegraphics{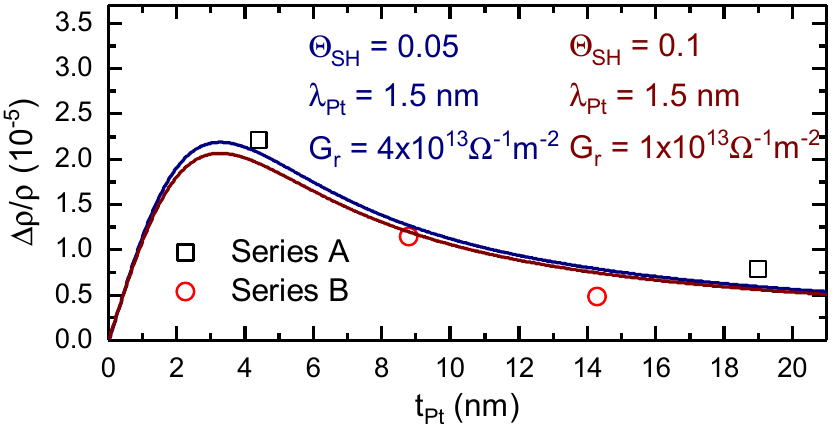}
    \caption{\label{fig:mrdpt}The magnitude of the SMR as a function of the Pt thickness is depicted here for the two sample sets. The established thickness dependence of the SMR is shown for two parameter sets as dark blue and dark red lines. The parameters adapted from Althammer \ea\ \cite{Althammer:2013} are summarized above the data.}
\end{figure}

In summary, we presented magnetoresistive measurements on YIG/Pt 
heterostructures, where the Pt is deposited via ALD. 
Our data suggest the presence of SMR and good
electrical properties of the Pt films that is comparable with sputtered films.
Therefore, we demonstrate the possibility of depositing high-quality Pt with ALD. This implies the technological feasibility of
3D conformal coating with spin-Hall-active materials, opening the door
to spin transport experiments in non-planar surface geometries.
However, because organic constituents are used in ALD precursors, further efforts to improve the YIG/Pt interface are necessary in order to obtain mixing conductance values that are comparable to platinum films deposited in ultra-high-vacuum.

We would like to thank S. Piontek and P. B{\"u}ttner for technical support, and we acknowledge financial support by the Deutsche Forschungsgemeinschaft via SPP 1538 (project no. GO 944/4 and TH 1399/5)

\bibliography{YIGALDPt.bib}

\end{document}